\documentstyle[12pt]{article}

\newcommand{\bt}{\beta}
\newcommand{\dlt}{\delta}
\newcommand{\Hi}{{\cal H}_i}
\newcommand{\Om}{\Omega}
\newcommand{\om}{\omega}
\newcommand{\vp}{\varphi}
\newcommand{\be}{\begin{equation}}
\newcommand{\ee}{\end{equation}}
\newcommand{\ra}{\rightarrow}
\newcommand{\lgl}{\langle}
\newcommand{\rgl}{\rangle}
\newcommand{\prt}{\partial}
\newcommand{\ep}{\varepsilon}
\newcommand{\dgr}{\dagger}
\newcommand{\sk}{\stackrel{\rightarrow}{k}}
\newcommand{\sr}{\stackrel{\rightarrow}{r}}

\begin{document}

\begin{center}
{\large{\bf Multichannel Approach to Clustering Matter} \\ [5mm]
V.I.Yukalov and E.P.Yukalova} \\ [3mm]
{\it International Centre of Condensed Matter Physics \\
Univesity of Brasilia, CP 04513\\
Brasilia, DF 70919--970, Brazil}
\end{center}

\vspace{12cm}



\newpage

\begin{abstract}

An approach is developed, combining the ideas of quantum statistical 
mechanics and multichannel theory of scattering, for treating statistical 
systems whose constituents can possess different bound states realized as 
compact clusters. The main principles for constructing multichannel 
cluster Hamiltonians are formulated: principle of statistical 
correctness, principle of cluster coexistence, and principle of 
potential scaling. The importance of the principle of statistical 
correctness is emphasized by showing that when it does not hold the 
behaviour of thermodynamic functions becomes essentially distorted. And 
moreover, unphysical instabilities can appear. The ideas are carefully 
illustrated by a statistical model of hot nuclear matter.

\end{abstract}

\newpage

\section{Introduction}

If the constituents of a statistical system strongly interact with each 
other, they can form different bound states exhibiting themselves as 
compact clusters. Ubiquitous chemical elements are examples of such 
clusters [1]. Another well known example is fog consisting of droplets 
made of water molecules. A variety of other examples are listed in ref.[2].

Despite their so widespread prevalence there is no good theory describing 
statistical clustering systems. It is possible to separate three problems 
in this description: One is the consideration of the dynamics of growing 
or evaporating clusters. This process can be portrayed  by complicated 
partial differential equations, similar to those that treat the dynamics 
of nuclei of a new phase or the growth and dissolution of macrodefects in 
nonequilibrium systems [3-5]. Another problem is the description of the 
properties of a single cluster inside equilibrium matter. This can be 
exemplified, for instance, by the model of a cluster in thermostate [2]. 
A fullerene molecule can also be treated as a large cluster in 
thermodynamically equilibrium surrounding [6,7]. A nucleon in nuclear 
matter can be considered as a quark bag or quark cluster [8,9]. The third 
problem is to give a statistical description of a quantum system whose 
particles can form various bound states, so that in the same system different
types of clusters arize.

In this paper we address the third problem of the mentioned three, that 
is, the statistical description of a quantum system with multiple 
clusters. Theoretical treatment of such a system is the least understood 
and developed, as copmared to the first two problems, There are several 
objective difficulties in treating clustering matter. The main of them 
are as follows.

Suppose that particles forming a statistical system have various bound 
states. Each bound state can be described by a Bethe--Salpeter--type 
equation for a $\;z\;$--particle Green function, where $\;z\;$ is the 
compositeness of the bound state. This type of equations, even for a 
two--particle Green function, is very difficult to deal with. When there 
are many bound states interacting with each other as well as with unbound 
particles, then one has to deal with a large system of many coupled 
Bothe--Salpeter equations for two--, three--, and so on up to the highest 
order $\;z\;$--particle Green functions, plus the Dyson--type equations for
single--particle Green functions [10]. Even to deal with a two--particle
Green function is not as easy. And imagine that one has to operate with a
ten--particle Green function, when there occur ten--particle bound states. 
It is evident that in the case of multiple bound states the standard approach 
becomes practically unsolvable and useless. Then one simplifies the 
problem by inventing effective thermodynamic potentials. However, because 
of ambiguity in constructing such potentials, the thermodynamic behaviour 
of the considered system may have quite different features depending on 
the assumptions used. Morover, the fabricated thermodynamic potential 
may have no relation not only to the considered clustering matter but 
even to any statistical system. Thus, the first question that arises is: 
What is a general criterion which any effective thermodynamical potential 
must satisfy for correctly representing a statistical system?

In order to simplify the problem one often postulates that solely one 
type of bound states can exist under the given thermodynamical 
parameters, say temperature and density. An ensemble of given one--type 
bound states then is equivalent to a thermodynamic phase. And a 
transformation of one phase into another goes through a genuine phase 
transition. Evidently, such a simplification is not merely too rough but 
contradicts the general properties of a quantum system in which bound 
and unbound states can be formed simultaneously. The latter is sinonymous 
to requiring that different types of clusters could coexist. The 
necessity of allowing for such a cluster coexistence is advocated and 
stressed in the present paper.

As soon as we accept the possibility of cluster coexistence, another 
problem, looking unsolvable, emerges. When the particles constituting a 
system can form several types of bound--state clusters, then it is 
necessary to define many interaction potentials between these clusters as 
well as between the clusters and the constituent particles. If the 
particles can make, say, ten types of clusters, then we need to know 
$\;C_{10}^2 + 10 =55\;$ different interaction potentials. Where the 
latter can be taken from?

The aim of this paper is twofold: To formulate the general principles of 
a correct statistical description of clustering matter and to illustrate 
these principles by examples.

\section{ Principle of Statistical Correctness}

Wishing to analyse thermodynamic properties of a statistical system, we 
need to define a thermodynamic potential. Writing the latter, one usually
invokes some simplifications leading to an effective thermodynamic 
potential $\;\Om(\vp)\;$ containing a set $\;\vp=\{\vp_i\}\;$ of 
auxiliary functions depending on space and/or thermodynamic variables. 
What necessary to keep in mind, first of all, is that not each effective
potential can have sense, however reasonable it may look. A thermodynamic
potential, to be accepted as such, must satisfy the properties formulated 
below.

\vspace{3mm}

{\it Property 1}. {\bf Statistical Representability}.

\vspace{2mm}

An effective thermodynamic potential $\;\Om(\vp)\;$ represents an equilibrium
statistical system if and only if it has the Gibbs form
\be
\Om(\vp) =\Om[H(\vp)]
\ee 
defined as
\be
\Om[H]\equiv -T\ln{\rm Tr}\;e^{-\bt H} ,
\ee
where $\;\bt\equiv T^{-1}\;$, and the dependence on auxiliary functions 
comes only through an effective Hamiltonian $\;H(\vp)\;$. Such a 
thermodynamic potential is called {\it statistically representable}.

\vspace{2mm}

Thus, if one invents an effective thermodynamic potential, even 
pronouncing seemingly plausible words, this does not necessary mean 
that the invented potential describes some statistical system. If this
 potential is not statistically representable it corresponds to no 
equilibrium statistical system. For example, a thermodynamic potential in 
the excluded--volume approximation is not statistically representable.

\vspace{3mm}

{\it Property 2}. {\bf Thermodynamic Equivalence}.

\vspace{2mm}

A statistical system defined by a Hamiltonian $\;H\;$ is thermodynamically
equivalent to a system modeled by an effective Hamiltonian $\;H(\vp)\;$ 
if and only if their thermodynamic potentials $\;\Om\;$ and $\;\Om(\vp)\;$ 
are statistically representable, i.e.
\be
\Om =\Om[H], \qquad \Om(\vp) =\Om[H(\vp)] ,
\ee
and are equal to each other,
\be
\Om[H] =\Om[H(\vp)] .
\ee
The corresponding Hamiltonians are called {\it thermodynamically equivalent}.

\vspace{2mm}

For the case of infinite matter, condition (4) can be weakened by requiring 
the validity of the asymptotic, in the thermodynamic limit, equality
$$ \lim_{V\ra\infty}\frac{1}{V}\left ( \Om[H] -\Om[H(\vp)]\right ) = 0 . $$

\vspace{3mm}

{\it Property 3}. {\bf Statistical Correctness}.

\vspace{2mm}

The necessary condition for an effective Hamiltonian $\;H(\vp)\;$ to 
be thermodynamically equivalent to a Hamiltonian $\;H\;$, not depending 
on auxiliary functions $\;\vp\;$, is the variational equality
\be
\left \lgl \frac{\dlt}{\dlt \vp} H(\vp)\right \rgl = 0 ,
\ee
where the variation over $\;\vp\;$ implies the set of variations with respect 
to each $\;\vp_i\;$, and the average of an operator $\;\hat A\;$ is
\be
\lgl\hat A\rgl \equiv 
\frac{{\rm Tr}\hat A\exp\{-\bt H(\vp)\}}{{\rm Tr}\exp\{-\bt H(\vp)\}} .
\ee
Equality (5) is an evident consequence of the property of thermodynamic
equivalence. We shall call (5) the condition of {\it statistical 
correctness}.

Basing on these properties, we formulate the 

\vspace{3mm}

{\bf Principle of Statistical Correctness}:

\vspace{1mm}

{\it An effective thermodynamical potential is statistically correct if it 
is statistically representable with an effective Hamiltonian satisfying 
the condition of statistical correctness}.

\vspace{2mm}

Suppose that the thermodynamic potential $\;\Om=\Om(\vp)\;$ is statistically
representable with the Hamiltonian
\be
H(\vp) =\hat E -\sum_{i}\mu_i\hat N_i ,
\ee
in which $\;\hat E\;$ is the energy operator, $\;\mu_i\;$ is a chemical
potential, and $\;\hat N_i\;$ is a number--of--particle operator. If this
thermodynamic potential is statistically correct, then condition (5) 
guarantees the validity of the thermodynamic relations
$$ p =-\frac{\prt\Om}{\prt V} =-\frac{\Om}{V} , $$
$$ \ep = T\frac{\prt p}{\prt T} - p +\sum_{i}\mu_i\rho_i =
\frac{1}{V}\lgl\hat E\rgl , $$
\be
s =\frac{\prt p}{\prt T} =
\frac{1}{T}\left (\ep + p -\sum_{i}\mu_i\rho_i\right ) ,
\ee
$$ \rho_i =\frac{\prt p}{\prt\mu_i} =\frac{1}{V}\lgl\hat N_i\rgl $$
for the pressure $\;p\;$, energy density $\;\ep\;$, entropy density $\;s\;$, 
and particle density $\;\rho_i\;$; the volume of the system being $\;V\;$.

For the effective thermodynamic potential not satisfying condition (5) 
the thermodynamic relations (8) would be broken, that is, the values of 
the quantities on the left--hand side of (8) calculated as the 
corresponding derivatives or as statistical averages would be different. 
Such an inconsistency in (8) would signify that one should not trust 
to predictions derived from an effective thermodynamic potential which 
is statistically incorrect.

The importance of sustaining the self--consistency in the thermodynamic
relations (8) for studying the thermodynamics of effective models was 
emphasized by Zim\'anyi et al. [11]. Requiring 
the validity of (8) for an effective thermodynamic potential containing 
unknown correcting functions yields a complicated system of nonlinear
differential equations in partial derivatives with respect to the 
variables $\;T,\; V\;$, and $\;\mu_i\;$. This system of equations is to 
be complimented by boundary conditions. Such a system has no unique solution. 
To extract somehow the latter, one needs several additional heuristic
assumptions and fitting parameters. Contrary to this, the principle 
of statistical correctness formulated above is, as we shall show in the 
next section, much simpler to deal with and gives a unique solution.

Also, the principle of statistical correctness is more general leading 
to (8) but not conversely. Moreover, if one finds correcting functions 
directly from the first--order relations (8), this does not guarantee 
the validity of the second order relations for the specific heat
\be
C_V =\frac{\prt\ep}{\prt T} =\frac{\bt^2}{V}\left (\lgl\hat E^2\rgl -
\lgl\hat E\rgl^2\right )
\ee
and the isothermic compressibility
\be
\kappa_T=-\frac{1}{V}\left (\frac{\prt p}{\prt V}\right )^{-1} =
\frac{\bt}{\rho^2V}\left (\lgl\hat N^2\rgl - \lgl\hat N\rgl^2\right ) .
\ee
At the same time, the condition of statistical correctness (5) insures 
that (8),(9),(10) and other analogous relations of arbitrary order hold true.

\section{Principle of Cluster Coexistence}

If the particles of a quantum system form different bound states, then 
the latter can influence thermodynamic properties. To take this influence 
into account, one has to allow for the existence of these bound states, 
estimating their relative contribution to the properties of the system. 
This is equivalent to saying that it is necessary to allow for the 
coexistence of different clusters, calculating their statistical weights 
in the analyzed properties. The necessity of taking into account both 
unbound and various bound states seems so natural from the general 
quantum--mechanical point of view that it could look excessive repeating 
it, if it would not be so common meeting in literature statistical models 
in which different cluster states are prohibited to coexist, being 
treated as pertaining to different thermodymanic phases. Because of this 
frequent confusion of quantum states with statistical states we feel it 
is worth clarifying this question once more. A quantum bound state 
corresponds to a cluster but not to a thermodynamic phase. Although it 
may happen that under some conditions, say at low temperature, one type 
of clusters prevails while at high temperature another type becomes 
dominant, this does not mean that there is no admixture of other types of 
clusters among those of a predominant type. Predominance of one type of 
clusters is not the same as the complete prohibition to exist for other 
types. What types of clusters and in which proportions can coexist in a 
particular thermodynamic phase is, figuratively speaking, to be decided 
by the system itself, which implies that each system tends to a state of 
maximal thermodynamic stability, and the concentrations of clusters are 
to be defined by stability conditions. The latter can be essentially 
spoiled by the prohibition for clusters to coexist and can lead to 
incorrect thermodynamic behaviour, for instance, to the appearance of 
spurious phase transitions or to the change of order of genuine phase 
transitions.

The most general and, probably, apparent way of understanding the 
structure of a Hamiltonian corresponding to a system of coexisting 
clusters is from the point of view of the multichannel theory of 
scattering [12] considering each type of bound states, i.e., of clusters, 
as a reaction channel of interacting particles.

The methods of the abstract multichannel theory of scattering may 
be applied to physical systems of different nature, in which the 
constituent particles can form various bound states. Below we 
briefly delineate the main general ideas that could be used for any 
kind of such systems.

Let a system be defined by a Hamiltonian $\;H\;$ which is a selfadjoint 
operator acting in the Hilbert space $\;{\cal H}\;$, called the space 
of quantum states. Assume that the constituent particles of the system 
can form bound states. Thus, in a quark--gluon system various hadron 
states can be formed. Enumerate all possible types of bound states by 
the index $\;i\;$, where $\;i=1\;$ stands for unbound particles. Each 
type of bound states is individualized by a set of corresponding characteristics, such as the compositeness number $\;z_i\;$ showing 
the number of bound particles, effective mass $\;m_i\;$, and a set of 
quantum numbers like spin, isospin, colour, baryon number, and so on. 
All quantum states associated with the same type of bound states, 
indexed by $\;i\;$, compose the subspace $\;{\cal H}_i\subset{\cal H}\;$. 
In other words, $\;{\cal H}_i=\hat P_i{\cal H}\;$ is a projection 
of $\;{\cal H}\;$. Generally, different quantum states can be made 
orthogonal to each other. This means that the subspaces $\;{\cal H}_i\;$ 
can be considered as mutually orthogonal, that is, for the reducing 
projections $\;\hat P_i\;$ one has $\;\hat P_i\hat P_j=\dlt_{ij}\;$. 
Such pairwise orthogonal projections $\;\hat P_i\;$ are called the 
channel projections, and the related subspaces $\;{\cal H}_i\;$ are termed 
the channels. A quantum state pertaining to the channel $\;{\cal H}_i\;$ 
is an $\;i\;$--channel state.

The set of channels $\;\Hi\;$ is complete if all possible $\;i\;$--channel
states span the whole space of quantum states $\;{\cal H}\;$. Then the 
latter is written as the direct sum $\;{\cal H}=\oplus_i\Hi\;$. This 
is equivalent to the resolution of unity $\;\hat 1 =\sum_{i}\hat P_i\;$.

Let the time evolution of quantum states in a channel $\;\Hi\;$ be defined 
by a selfadjoint operator $\;H_i\;$. This implies that the channel 
$\;\Hi\;$ is invariant under the action of $\;H_i\;$. Because of the 
pairwise orthogonality of the channels, the operators $\;H_i\;$ are 
pairwise commuting, $\;[H_i,H_j]=0\;$. Each of such operators $\;H_i\;$ 
is called the channel Hamiltonian. The channel system is a system 
$\;\{ H_i\}\;$ of the channel Hamiltonians, each of which acts in 
the corresponding channel $\;\Hi\;$. The sum $\;\sum_{i}H_i+const\;\hat 1\;$ 
of the channel Hamiltonians, attributed to the channel system, may be 
named the multichannel Hamiltonian. The latter, by construction, acts 
in the space of quantum states $\;\oplus_i\Hi\;$.

In this way, the description of an ensemble of particles forming bound 
states can be done by constructing the corresponding channel system. 
To this end, one has to classify different types of states and to define 
the related channel Hamiltonians $\;H_i\;$. The sum 
$\;\sum_{i}H_i+const\;\hat 1\;$ gives the Hamiltonian of the channel 
system. It is worth noting that the multichannel Hamiltonian does not 
necessarily coincide with the exact Hamiltonian $\;H\;$, but rather 
gives a physically reasonable first approximation. In particular 
calculations, one can, if necessary, resort to perturbation theory 
starting from the multichannel Hamiltonian. Though in many applications 
already the channel approximation yields quite accurate results.

Summarizing what is said above we formulate the

\vspace{3mm}

{\bf Principle of Cluster Coexistence:}

\vspace{2mm}

{\it If the particles of an equilibrium statistical system can form 
clusters of different types, then all such cluster types can coexist with 
the probabilities defined by the condition of thermodynamic stability}.

\vspace{2mm}

Let us concretize how one can define such channel probabilities. 
The multichannel Hamiltonian, in general, reads
\be
H = \sum_i H_i +CV ,
\ee
where $\;H_i\;$ is an $\;i\;$--channel Hamiltonian and $\;CV\;$, a 
nonoperator term.

Take the channel hamiltonians, $\;H_i\;$, in the mean--field approximation
\be
H_i =\sum_k\om_i(k)a_i^\dgr(\sk)a_i(\sk)
\ee with an effective spectrum
$$ \om_i(k) =K_i(k) +U_i -\mu_i , $$
in which $\;k\;$ is the absolute value of the momentum $\;\sk\;$; 
$\;K_i(k)\;$ is the kinetic--energy term; $\;U_i\;$, a mean field; and 
$\;\mu_i\;$, the chemical potential of the $\;i\;$--type cluster. The 
field operator $\;a_i(\sk)\;$ is a column in the space of internal 
degrees of freedom, such as spin, flavor, and so on. For the mean--field
Hamiltonian (12) the momentum distribution
$$ \lgl a_i^\dgr(\sk)a_i(\sk)\rgl =\zeta_in_i(k) $$
is easily calculated; $\;\zeta_i\;$ being a degeneracy factor, and
$$ n_i(k) \equiv [\exp\{\beta\om_i(k)\mp 1\}]^{-1} $$
is the Bose-- or Fermi function depending on the upper or lower sign, 
respectively. The average density of the $\;i\;$--type clusters is
\be
\rho_i =\frac{\zeta_i}{(2\pi)^3}\int n_i(k)d\sk .
\ee
Note that the isotropicity of the system is assumed here, because of 
which $\;n_i(k)\;$ depends on $\;k\equiv|\sk|\;$. Consequently, in (13) 
we could write
$$ \int n_i(k)d\sk = 4\pi\int_0^\infty n_i(k)k^2dk ; $$
but for the sake of brevity we prefer the former notation.

The density of $\;i\;$--clusters (13) factored by the compositeness 
number $\;z_i\;$ gives the density of particles $\;z_i\rho_i\;$ that are 
bound in the clusters of the $\;i\;$--type. The total average density of 
particles is
\be
\rho =\sum_i z_i\rho_i .
\ee The statistical weight of each channel is characterized by the {\it 
channel probability}
\be
w_i \equiv z_i\frac{\rho_i}{\rho} ,
\ee 
of the $\;i\;$--type clusters. By definition, eq. (15) enjoys the conditions
\be
0\leq w_i\leq 1 , \qquad \sum_i w_i = 1 .
\ee
The compositeness number of unbound particles, $\;z_i\;$, is assumed to 
equal one.

The chemical potentials of $\;i\;$--clusters, entering into (15), can be 
expressed through the given thermodynamic variables with the use of the 
equilibrium conditions considered. Thus, if the average density of 
particles (14) is given, then the $\;i\;$--cluster chemical potentials 
$\;\mu_i\;$ are related to that of unbound particles, $\;\mu\;$, by the 
equilibrium condition
\be
\frac{\mu_i}{z_i} =\mu \qquad (\rho = const) .
\ee
In the case when the average baryon density
\be
n_B = \sum_i B_i\rho_i
\ee
is conserved, where $\;B_i\;$ is the baryon number of a cluster in an 
$\;i\;$--channel, then the equilibrium condition reads
\be
\frac{\mu_i}{B_i} =\mu_B \qquad (n_B=const) ,
\ee
where $\;\mu_B\;$ is the baryon potential of matter.

The cluster probabilities in (15), together with the kind of equilibrium 
considered, are defined as functions of thermodynamic variables. Finally, 
it is necessary to check that the considered equilibrium is stable, so 
that the stability conditions
\be
C_V > 0 , \qquad \kappa_T > 0
\ee
for the specific heat (9) and isothermic compressibility (10) are fulfilled.

In this way, all cluster probabilities in (15) can be calculated 
self--consistently, thus showing the proportions in which various types 
of clusters are intermixed in matter.

\section{Principle of Potential Scaling} 

In order to complete the definition of a system containing different 
types of clusters, we need to specify the interaction potentials 
$\;\Phi_{ij}(r)\;$ of these clusters. If there is a number of such 
cluster types, then we have to define numerous potentials $\;\Phi_{ij}(r)\;$. 
For example, as is mentioned in Introduction, in the case of $\;10\;$ 
cluster types we need to have $\;55\;$ such potentials. And for $\;100\;$ 
cluster types we would need already $\;5050\;$ interaction potentials. 
Where could we take them from?

The idea of how the interaction potentials could be connected with each 
other comes from the form of the equilibrium conditions (17) and (19). By 
analogy to these conditions we may conjecture that the interaction 
potentials are scaled, by means of the corresponding compositeness numbers,
to a universal function as follows:
\be
\frac{\Phi_{ij}(r)}{z_iz_j} =\Phi(r) .
\ee
Such a relation, as is clear, can be sensible only if the interaction 
potentials are of similar nature. To be more precise, we shall say that 
the interaction potentials are in the same {\it universality class} if 
and only is they can be scaled to one universal function, as in (21). As 
a counterexample we may adduce a pair of potentials one of which 
decreases, as $\;r\ra\infty\;$, and another increases. Certainly, these 
potentials cannot be in the same universality class. Only the clusters of 
similar nature possess the interaction potentials that can pertain to one 
such a class.

Combining what is said above, we come to the

\vspace{3mm}

{\bf Principle of Potential Scaling:}

\vspace{2mm}

{\it The clusters of similar nature interact with each other through 
potentials from the universality class, so that these potentials are 
scaled by the corresponding compositeness numbers to a universal function}.

\vspace{2mm}

The scaling relation (21), defining a universality class, can be derived 
from the following reasoning. Consider the channel reaction
\be
m+n+j\ra i+j
\ee
with the corresponding equation for the compositeness numbers
\be
z_m+z_n+z_j = z_i + z_j .
\ee
The reaction (22), with eq. (23), signifies that the $\;m\;$--cluster and 
$\;n\;$--cluster fuse together in the presence of a $\;j\;$--cluster. 
Assume the additivity of interactions with respect to the 
$\;j\;$--channel, that is,
\be
\Phi_{mj}(r) +\Phi_{nj}(r) =\Phi_{ij}(r) .
\ee
Then, from (23) and (24) it is straightforward to derive (21). Therefore, 
the scaling relation (21) can be interpreted as the result of the 
additivity of interactions (24) under condition (23). Making this 
statement more general, but less accurate, we may say that the scaling 
relation (21) is a manifestation of two conservation laws: conservation 
of energy and conservation of particle number during the reactions of 
fusion and decay.

Thus, if at least one of the interaction potentials is known, say 
$\;\Phi_{mn}(r)\;$, then by means of the scaling relation (21), all 
others can be expressed through this reference potentials as
\be
\Phi_{ij}(r) =\frac{z_iz_j}{z_mz_n}\Phi_{mn}(r) .
\ee
This solves the problem of multiple interaction potentials making the 
multichannel approach to clustering matter completely defined.

\section{Clustering Nuclear Matter}

To illustrate the approach, we need to apply it to a particular 
clustering substance. it is possible to distinguish two quite different 
cases: One, when the consideration can be limited by several cluster 
types, say, by about ten of them; and another, when there exists a 
multitude of various cluster types. The latter case has to do with 
substances like fog which consists of droplets whose compositeness 
numbers range from one to many billions. A good example of the former 
case is nuclear matter which can be characterized by a limited number of 
clusters. In this way, fog is a more complicated system, which we shall 
consider in a separate publication. And here we analyse the more simple 
case of hot and dense nuclear matter, paying the main attention to the 
process of evaporation and condensation of hadron clusters, that is, to 
the deconfinement--confinement transition.

The necessity of resorting to statistical models for nuclear 
quark--hadron matter is due to the fact that perturbative quantum 
chromodynamics  does not provide information on the equation of state in 
the whole region of thermodynamic parameters. Especially little can be 
said about the most interesting region of deconfinement--confinement 
transition. There exist lattice calculations which, however, are reliable 
only for the case of zero baryon density, but not for its finite values [13].

The overwhelming majiority of statistical models for deconfinement have 
been based on the assumption that the latter is a phase transition 
between two pure phases, the clustered hadron phase and the phase of 
unbound quarks and gluons called the quark--gluon plasma. But this assumption 
contradicts to many computer simulations and also to some analytical 
estimates. For example, intensive numerical simulations for lattice 
quark--gluon plasma [14] and pure gluon plasma [15] revealed nontrivial 
effects due to strong particle correlations at temperatures above the 
deconfinement temperature $\;T_d\;$, which has been interpreted as the 
existence of hadronic modes, even at $\;T > T_d\;$. The deconfinement was 
found to be a rapid crossover, but not a genuine phase transition [14]. 
The high--temperature phase contains fluctuations being colour singlet 
modes, hadronic in character. This is because the poles and cuts in the 
linear responce functions of the hadronic phase go over smoothly into those
of the high--temperature one, so that all low--temperature hadrons have 
analogs in the high--temperature phase. The low--and high--temperature 
phases both have confining characteristics, but the effect of confinement 
upon thermodynamic properties becomes less and less significant as 
temperature increases.

The confinement--deconfinement transition can be compared with the 
insulator--metal transition or with the ionization, since hadrons are 
nothing but the bound states of quarks and gluons, or small droplets of 
quark--gluon plasma [16-18]. In insulating solids, below the transition 
point to a metal, the conductivity is not strictly zero, since the 
ionization energy is finite and ionization can locally provide some free 
electrons. Similarly, in the hadron matter below the deconfinement 
temperature quarks can be separated. Both the conventional Mott 
transition in solids and the deconfinement transition in hadronic matter 
thus lead from a regime, in which the binding can locally be broken by 
ionization, to one where it is globally removed by a collective screening 
of the binding force.

The spatial structure of correlation functions, obtained in lattice 
numerical simulations for hot quark--gluon plasma, is very similar to the 
structure of the corresponding zero--temperature functions [14]. The 
thermodynamic characteristics, such as pressure and energy density, aslo 
display strong nonperturbative effects persisting till about $\;2T_d\;$, 
as is found in the lattice simulations and discussed in recent surveys 
[19,20]. The fact that some lattice simulations with dynamical quarks 
displayed jumps of thermodynamic characteristics at  $\;T_d\;$ can be 
explained [21] as merely due to the finite size of the lattice, since the 
transition becomes less and less abrupt as the lattice size inreases. 
Lengthy runs show no evidence for metastability thus suggesting that 
there is no sharp transition, but only a crossover phenomenon.

The coexistence scenario, based on numerical simulations, is supported 
also by some analytical calculations. For instance, the density--density 
correlation functions in the Nambu--Jona--Lasinio model contain poles and 
cuts that are the same at all temperatures [22]. Therefore meson modes do 
exist above as well as below transition temperature. Analogously, quarks 
and gluons should also exist in the low--temperature as well as in 
high--temperature phase. The graduate change in the excitation spectrum 
from hadronic states to quarks and gluons and the survival of hadronic 
modes in the high--temperature phase appear also in the magnetic--current 
approximation [23] and in the instanton--liquid approach [24]. This 
suggests the following picture of the quark--hadron coexistence. The 
correlations between quarks persist above the deconfinement temperature 
forcing some of them to correlate into colour singlets. As the quarks are 
moving in the heat bath, the strings connecting them for colour 
neutrality are constantly breaking and reforming, which can be 
interpreted as hadrons going in and out of the heat bath. This picture 
has a formal resemblance to the string--flip model [25] although with 
time--like strings.

Thus, the concept of cluster coexistence is absolutely natural from the 
point of veiw of the multichannel scattering theory and is supported by 
lattice numerical calculations. The cluster coexistence is somewhat 
similar, although not identical to heterophase coexistence common for 
many statistical systems, both these types of coexistence leading to 
precursor, or pretransitional, phenomena [26].

There is a variety of different statistical models trying to describe 
deconfinement in nuclear matter (see reviews [27,28]). The majority of 
these models, with a few exceptions (e.g. [29-31]), do not take into 
account the possible coexistence of clusters. This is why the predictions 
of such models are in disagreement with lattice numerical results.

Consider the clustering nuclear matter whose elementary constituents are 
quarks, antiquarks and gluons. These can form bound states corresponding 
to hadron clusters. The total set $\;\{ i\}\;$ of the indices enumerating 
the channels can be separated into two groups, $\{ i\}_1\;$ and 
$\;\{ i\}_z\;$. The first group $\;\{ i\}_1\;$ is related to unbound
constituents, quarks, antiquarks, and gluons whose compositeness number
$\;z_i=1\;$. The second group corresponds to hadron channels representing 
bound states with compositeness number $\;z_i\geq 2\;$. Respectively, the 
density (14) may be written as the sum
$$ \rho=\rho_1 +\rho_z , \qquad \rho_1\equiv \sum_{\{ i\}_1}\rho_i ,
\qquad \rho_z\equiv \sum_{\{ i\}_z} z_i\rho_i , $$
in which $\;\rho_1\;$ is the density of unbound particles and 
$\;\rho_z\;$ is the density of particles in bound states.

Accept the effective Hamiltonian (11) with the channel Hamiltonians given 
by (12). To define the effective spectrum $\;\om_i(k)\;$ in (12), we have 
to concretize the mean field $\;U_i\;$. The quark--gluon plasma mean 
field $\;U_1\;$ can be defined as
\be
U_1\equiv U(\rho) =\rho\int V(r)s(r)d\sr ,
\ee
where $\;V(r)\;$ is a confining potential and $\;s(r)\;$, a screening 
function. The necessity of including a screening correlation function 
into the mean--field approximation is the general requirement for any 
statistical system with nonintegrable interaction potentials [32]. The 
approximation (26) is the correlated Hartree approximation. The confining 
potential is not integrable since its mostly often used representation 
has the power--law behaviour
\be
V(r)=Ar^{\nu} \qquad (0<\nu\leq 2) .
\ee
The screening function
$$ s(r) =\bar s\left (\frac{r}{a}\right ) \qquad (a\equiv\rho^{-1/3}) $$
can be scaled by the mean interparticle distance $\;a\;$. Then, with the 
notation
\be
J^{1+\nu} \equiv 4\pi A\int_0^\infty\bar s(x)x^{2+\nu}dx
\ee
for the effective intensity of interactions $\;J\;$ measured in energy 
units, the mean field (26) becomes
\be
U(\rho)=J^{1+\nu}\rho^{-\nu/3} .
\ee
The plasma mean field (29) has the following asymptotic properties:
\begin{eqnarray}
U(\rho) \ra\left\{\begin{array}{cc}
\infty , & \rho\ra 0 , \\
\\
0 ,      & \rho\ra\infty .\end{array}\right.
\end{eqnarray}
The upper line here tells that quarks and gluons cannot exist as free 
particles outside dense nuclear matter -- this is what is called the 
colour confinement. The lower line in (30) shows that the interparticle
interaction decreases with the decrease of the average distance between 
particles -- this is the so called phenomenon of asymptotic freedom.

The mean field for an $\;i\;$--channel corresponding to a bound hadron 
state can be written as
\be
U_i =\sum_{\{ j\}_z}\Phi_{ij}\rho_j + z_i[U(\rho) - U(\rho_z)] ,
\ee
where the summation is over the bound states, and $\;\rho_z\;$ is the 
density of particles in bound states. The first term in (31) desribes the 
interaction of a cluster in an $\;i\;$--channel with all other hadrons; 
the mean interaction potential being
$$ \Phi_{ij} =\int\Phi_{ij}(r)d\sr , \qquad 
\Phi_{ij}(r) \equiv V_{ij}(r)s_{ij}(r) , $$
where $\;V_{ij}(r)\;$ is a bare interaction potentials between the clusters 
of the $\;i\;$-- and $\;j\;$--types, and $\;s_{ij}(r)\;$ is a screening 
correlation function, so that $\;\Phi_{ij}(r)\;$ is called the screened, or
effective, potential. The second term in (31), in the square brackets, 
models the interaction of a cluster of the $\;i\;$--channel with unbound 
plasma states.

Invoking the principle of potential scaling (21) for the effective 
potentials $\;\Phi_{ij}(r)\;$, we may write
\be
\Phi_{ij} = z_iz_j\Phi , \qquad 
\Phi\equiv \int\Phi(r)d\sr , 
\ee
where $\;\Phi(r)\;$ is some reference function. Then (31) reduces to
\be
U_i =z_i\Phi\rho_z + z_iJ^{1+\nu}\left ( \rho^{-\nu/3} -\rho_z^{-\nu/3}
\right ) .
\ee

At this point it is necessary to stress that the interaction between 
unbound particles and these between hadron clusters are of different 
nature, the former growing while the latter diminishing with increasing 
interparticle distance. This is equivalent to saying that they are from 
different universality classes, thus, cannot be scaled to the same 
reference function. Therefore, the reference function $\;\Phi(r)\;$ in 
(32) cannot be interpreted as an interaction potential for unbound plasma 
constituents. Consequently, with these two universality classes, we have 
to keep two parameters, $\;\Phi\;$ and $\;J\;$, which are to be chosen 
independently.

The Hamiltonian (11) must satisfy the principle of statistical 
correctness (see Sec.2). The role of auxiliary functions in (11) is 
played by the densities $\;\rho\;$ and $\;\rho_z\;$, which are functions 
of temperature and baryon density. Therefore, the condition of 
statistical correctness (5) reads
\be
\left \lgl\frac{\delta H}{\delta\rho}\right\rgl = 0 , \qquad
\left\lgl\frac{\delta H}{\delta\rho_z}\right\rgl = 0 .
\ee
Note that instead of $\;\rho\;$ and $\;\rho_z\;$, as auxiliary functions, 
we could take $\;\rho\;$ and $\;\rho_1\;$ or $\;\rho_1\;$ and $\;\rho_z\;$. 
The condition (34) does not depend on this choice, since 
$\;\rho=\rho_1+\rho_z\;$. 
Substituting (11) and (12) into (34) gives
$$ \sum_i\rho_i\frac{\delta U_i}{\delta\rho} +
\frac{\delta C}{\delta\rho} = 0 ,  \qquad
\sum_i\rho_i\frac{\delta U_i}{\delta\rho_z} +
\frac{\delta C}{\delta\rho_z} = 0 . $$
Taking account of (31) and (33) yields 
$$ \frac{\delta C}{\delta\rho} =\frac{\nu}{3}J^{1+\nu}\rho^{-\nu/3} ,
\qquad \frac{\delta C}{\delta\rho_z} = -
\frac{\nu}{3}J^{1+\nu}\rho_z^{-\nu/3} -\Phi\rho_z . $$
The solution of these equations is straightforward and, up to a constant, 
it is
\be
C =\frac{\nu}{3-\nu}J^{1+\nu} \left ( \rho^{1-\nu/3} -\rho_z^{1-\nu/3}\right )
-\frac{1}{2}\Phi\rho_z^2 .
\ee

In this way, the multichannel Hamiltonian (11) is completely defined. 
Emphasize that the term $\;CV\;$ cannot be omitted since its presence 
provides the validity of the principal of statistical correctness. Only 
retaining this term makes it possible to find the correct thermodynamic 
behaviour of the clustering matter.

With the Hamiltonian (11), the pressure is
\be
p =\sum_ip_i , \qquad
p_i =\pm T\frac{\zeta_i}{(2\pi)^3}\int\ln\left [1\pm n_i(k)\right ]d\sk ,
\ee
and for the energy density one has 
\be
\ep =\sum_i\ep_i , \qquad
\ep_i = \frac{\zeta_i}{(2\pi)^3}\int\left [ \om_i(k)+\mu\right ] n_i(k)d\sk .
\ee
Taking the kinetic--energy term in the relativistic form 
$\;K_i(k)=\sqrt{k^2+m_i^2}\;$, where $\;m_i\;$ is the corresponding mass, 
we get the spectrum
\be
\om_i(k) =\sqrt{k^2+m_i^2} + U_i -\mu_i .
\ee

Before analysing in detail the thermodynamic behaviour of the system, 
let us notice that it has the following properties. In the case when 
$\;T\ra 0\;$ and $\;n_B\ra 0\;$, all particles are condensed into hadron 
clusters. And when $\;T\ra\infty\;$, at any fixed $\;n_B\;$, only unbound 
states survive. Consider the high--temperature case more accurately, 
since in this limit we may compare the results with the available 
perturbative calculations in quantum chromodynamics.

When $\;T\ra\infty\;$, then only quarks, antiquarks, and gluons remain, 
so that the set of unbound states $\;\{ i\}_1\;$ is $\;\{ q,\bar 
q,g\}\;$. The pressure (36) and energy density (37) become
\be
p\simeq \sum_{\{ i\}_1} p_i , \qquad \ep \simeq \sum_{\{ i\}_1}\ep_i .
\ee
At temperatures much higher than the quark masses the latter can be 
neglected, because of which in (36) and (37) we may substitute
\be 
n_i(k)\ra \left [\exp\{\beta (k-\mu_i)\}\mp\right ]^{-1} ,
\ee
the chemical potentials being
\be
\mu_q =-\mu_{\bar q}\equiv\mu , \qquad \mu_g = 0.
\ee
Thus we come to
\be
p_i=\frac{\zeta_i}{6\pi^2}\int_0^\infty
\frac{k^3dk}{\exp\{\beta (k-\mu_i)\}\mp 1} , \qquad \ep_i =3p_i .
\ee
Take into account that the degeneracy factors of quarks and antiquarks 
are the same, $\;\zeta_g=\zeta_{\bar q}\;$, and that these particles are 
fermions while gluons are bosons. Then an exact integration yields
$$ p_q +p_{\bar q} =\frac{\zeta_q}{12}\left (
\frac{7\pi^2}{30}T^4 +\mu^2T^2 +\frac{\mu^4}{2\pi^2}\right ) , \qquad
p_g =\frac{\pi^2}{90}\zeta_gT^4 . $$
From here, for the pressure in (39) we have
\be
p\simeq 
\frac{\pi^2}{90}\left (\frac{7}{4}\zeta_q +\zeta_g\right ) T^4 
+\frac{\zeta_q}{12}\mu^2T^2\left ( 1 +\frac{\mu^2}{2\pi^2T^2}\right ) . 
\ee
For the specific heat we find
\be
C_V \simeq \frac{2\pi^2}{15}\left (\frac{7}{4}\zeta_q +
\zeta_g\right ) T^3 + \frac{\mu^2(\mu^2-\pi^2T^2)}{2(3\mu^2+\pi^2T^2)}
\zeta_qT^2 .
\ee

The pressure (43) is to be compared with that obtained by perturbation 
theory in quantum chromodynamics (see Appendix). Due to the relations
\be
\zeta_q = 2\times N_f\times N_c , \qquad \zeta_g =2\times (N_c^2 -1 )
\ee
for the degeneracy factors of quarks and gluons, where $\;N_f\;$ and 
$\;N_c\;$ are the numbers of flavours and colours, respectively, we make 
it sure that the high--temperature pressure (43) asymptotically coincides 
with the perturbative pressure in $\;QCD\;$ , as $\;t\ra\infty\;$.

Expression (43) can be simplfied further if we substitute there the 
chemical potential as a function of $\;T\;$ and $\;n_B\;$, which can be 
found from the formula for the baryon density
\be
n_B\simeq \frac{1}{3}(\rho_q -\rho_{\bar q}) =
\frac{\zeta_q}{3(2\pi)^3}\int [n_q(k) - n_{\bar q}(k)]d\sk .
\ee
With eq.(40), this gives
\be
n_B\simeq \frac{\zeta_q\mu}{18\pi^2} (\mu^2 +\pi^2T^2) .
\ee
From (47), we get the chemical potential
\be
\mu\simeq \frac{18n_B}{\zeta_qT^2} \qquad (T\ra\infty) .
\ee
For the density of quarks and gluons we find
$$ \rho_q \simeq \frac{3\zeta_q}{4\pi^2}\zeta(3)T^3 , \qquad
\rho_g \simeq \frac{\zeta_g}{\pi^2}\zeta(3)T^3 , $$
where $\;\zeta(3)=1.20206\;$. Using (48) and introducing the notation
\be
\zeta\equiv \frac{7}{4}\zeta_q +\zeta_g ,
\ee
we come to the conclusion that the high--temperature behaviour of the 
system asymptotically reduces to that of the Stephan--Boltzmann plasma,
$$ p\simeq p_{SB} , \qquad \ep\simeq\ep_{SB} , \qquad C_V\simeq C_{SB} , $$
for which
\be
p_{SB} =\frac{\pi^2}{90}\zeta T^4 , \qquad \ep_{SB} 
=\frac{\pi^2}{30}\zeta T^4 , \qquad C_{SB} =\frac{2\pi^2}{15}\zeta T^3 ,
\ee
with the factor (49). In what follows it will be convenient to present 
the results in a dimensionless form scaling them by means of the 
Stephan--Boltzmann expressions in (50). For the number of colours $\;N_c 
=3\;$, we have $\;\zeta_q=6N_f\;$ and $\;\zeta_g=16\;$. Therefore, the 
factor (49) is
$$ \zeta =16\left ( 1 +\frac{21}{32}N_f\right ) , $$
which is to be substituted into (50). Then, e.g., the pressure becomes
$$ p_{SB} =\frac{8\pi^2}{45}\left ( 1 +\frac{21}{32} N_f\right ) T^4 . $$

\section{Analysis of Thermodynamic Characteristics} 

To accomplish explicit calculations of thermodynamic characteristics, we 
need to specify three parameters, $\;J,\;\Phi\;$, and $\;\nu\;$. The first 
of them, $\;J\;$, describes the intensity of nonperturbative interactions 
in the quark--gluon plasma. We can evaluate $\;J\;$ in several ways. 
The simplest way is to remember that nonperturbative effects in the 
quark--gluon plasma are commonly associated with the so--called bag 
constant, $\;B\;$, for which one uses different values so that 
$\;B^{1/4}\;$ lies in the interval between about $\;150\;$ to $\;300\;MeV\;$. 
We can accept for $\;J\;$ the value from the middle of this interval, 
that is $\;J=225\;MeV\;$.

The constant $\;\Phi\;$, according to (32), can be chosen as 
$\;\Phi =\Phi_{33}/9\;$ with $\;\Phi_{33}\;$ given by 
$\;\Phi_{33}=\int V_{33}(r)s_{33}(r)dr\;$, where 
$\;V_{33}(r)\;$ is the interaction potential between the three--quark 
bound states, that is between nucleons. The nucleon interaction 
potentials are well known from scattering experiments. There are 
several representations for these potentials. For nonintegrable hard--core
potentials one has to take into account the correlation function
$\;s_{33}(r)\;$, while for integrable soft 
potentials $\;s_{33}(r)\approx 1\;$. Among many known nucleon--nucleon
potentials, we prefer the Bonn potential [33], which provides an 
accurate description of nucleon scattering and has sufficiently simple 
analytic form. Following the common consensus that thermodynamics of 
nuclear matter does not depend on the mutual orientation of spins of 
scattering nucleons, we average over spin directions, and assume also 
that the interaction between any pair of nucleons, whether these are 
protons or neutrons, is the same. The so--called cut--off terms of the 
Bonn potential can be neglected since they start playing an essential 
role only for very short distances $\;\leq 0.1\;fm\;$, which would 
correspond to the baryon density $\;n_B\geq 10^3n_{0B}\;$ at which it 
would be hard to expect that any nucleons could survive. Here and in 
what follows $\;n_{0B}=0.167\;fm^{-3}\;$ is the normal baryon density. 
The parameter $\;\Phi\;$ calculated in this way is $\;\Phi=35\;MeV\;fm^3\;$.

The constant $\;\nu\;$ characterizes the power--law behaviour of the 
confining potential (27). The most frequently accepted cases are those 
of linear and harmonic confinement. Actually, our results do not change
qualitatively for $\;\nu\;$ between these two possibilities. Let us 
consider, for example, the harmonic confinement. As we have checked, 
our results practically do not change quantitatively in the region
$\;1.5\leq\nu\leq 2\;$. Thus, in what follows $\;\nu\approx 2\;$.

Finally, we have to concretize the reaction channels that will be included 
into consideration. In principle, the developed approach permits us to 
include any number of particles. We have analysed many variants which, 
as we found, demonstrate similar behaviour. Not to overload this paper, 
here we limit ourselves by the particles listed in the Table. We included 
into consideration multiquark states, although their status is not 
yet absolutely clear, because there has been an intensive discussion 
about their possible presence in nuclei (see reviews [34-36]) motivated 
by the Baldin interpretation of the commulative effect [37]. The chosen 
mass of the $\;6\;$--quark state corresponds to an average over the 
masses of several light dibaryons that are claimed to be observed in 
experiments [38]. The $\;9\;$-- and $\;12\;$--quark parameters are 
elicited from the bag model calculations [39].

We have analysed in detail the behaviour of the main thermodynamic
characteristics of the model as functions of temperature $\;\Theta=k_BT\;$ 
and relative baryon density $\;n_B/n_{0B}\;$. Fig.1 shows the probability 
of the plasma channel
$$ w_{pl} =\frac{1}{\rho}\left ( \rho_g + \rho_u + \rho_{\bar u}
+\rho_d + \rho_{\bar d}\right ) . $$
The probabilities of hadron channels are displayed in the following figures: 
in Fig.2, the pion--channel probability
$$ w_\pi =
\frac{2}{\rho}\left (\rho_{\pi^+} +\rho_{\pi^-} +\rho_{\pi^0} \right ) , $$
in Fig.3, the summarized, excluding pions, probability of other meson 
channels
$$ w_{\eta\rho\om} =\frac{2}{\rho}\left (
\rho_\eta + \rho_{\rho^+} + \rho_{\rho^-} +\rho_{\rho^0} +\rho_\om\right ) , $$
in Fig.4, the nucleon--channel probability
$$ w_3 =\frac{3}{\rho}\left ( \rho_n +\rho_{\bar n} +\rho_p +\rho_{\bar p}
\right ) , $$
in Fig.5, the probability of the six--quark channel
$$ w_6 =\frac{6}{\rho}\left ( \rho_{6q} + \rho_{6\bar q}\right ) , $$
and in Fig.6, the probability of the six--quarks in the Bose--condensed 
state. The probabilities of heavier multiquark channels are not depicted
since they are negligibly small,
$$ w_9 = \frac{9}{\rho}\left (\rho_{9q} + \rho_{9\bar q}\right ) < 10^{-3} , 
\qquad w_{12} = 
\frac{12}{\rho}\left (\rho_{12q} + \rho_{12\bar q}\right ) < 10^{-5} . $$
Figs.7--9 present some channel probabilities at zero temperature for a 
larger region of the relative baryon density. As is seen, the 
channel probabilities are continuous functions of their variables. 
The location of the deconfinement transition can be conventionally 
associated with the maxima of some of the channel probabilities.

Fig.10 demonstrates the pressure
$$ p = \sum_{i}p_i - C, \qquad  p_i = 
\pm T\frac{\zeta_i}{(2\pi)^3}\int\ln\left [ 1 \pm n_i(k)\right ] d\sk ; $$
and Fig.11, the energy density
$$ \ep = \sum_{i}\ep_i + C , \qquad
\ep_i = \frac{\zeta_i}{(2\pi)^3}\int\om_i(k)n_i(k)d\sk +
B_i\mu_B\rho_i . $$ 
These functions are quite monotonous but their ratio
$$ \frac{p}{\ep} = c_{eff}^2 , $$
having the meaning of the effective sound velocity squared, exhibits in 
Fig.12 a maximum around $\;T_d =160\;MeV\;$. The temperature dependence 
of $\;c_{eff}^2\;$ at $\;n_B=0\;$ agrees with that reconstructed from 
lattice data.

The specific heat in Fig.13 is also monotonous, but the reduced specific 
heat
$$ \sigma_V = \frac{T}{\ep}\frac{\prt\ep}{\prt T} $$
in Fig.14 again displays the maximum that can be associated with the 
location of the deconfinement line. The similar maximum exists for 
the compression modulus
$$ \kappa_T^{-1} = n_B\frac{\prt p}{\prt n_B} $$ 
in Fig.15.

These thermodynamic characteristics demonstrate that the transition from 
bound hadron states to unbound quark--gluon states is a gradual crossover. 
All these states are different channels of the same quantum system. Each 
channel is a subspace of the total Hilbert space of quantum states. 
Therefore, all channels, as possible states, always exist. But their 
statistical weights, defined by the channel probabilities, are different 
and change with varying thermodynamic parameters such as temperature and 
baryon density. The channel probabilities depend also on the Hamiltonian
parameters, for instance, masses. Thus, if we take the mass of a 
six--quark state $\;m_6 >2000\; MeV\;$, then the probability of this 
channel drastically drops down. Generally, among the members of the 
same family of states only those with lower masses are the most important. 
The states with heavier masses have, as a rule, much smaller statistical
weights.

Note also that the temperature behaviour of thermodynamic characteristics 
at nonzero baryon density is modified as compared to that at zero 
baryon density: Increasing $\;n_B\;$ smoothes the transition. Therefore 
one has to be quite cautious about those predictions for the processes 
at nonzero baryon density, which have been made basing on the processes 
at $\;n_B=0\;$.

In conclusion, we would like to emphasize that the main in this paper 
are not particular numerical predictions, though, we think, they are 
quite reasonable, but rather the whole picture and the ideology which 
the latter is based on. The basis of the model suggested rests upon 
the multichannel interpretation of different bound states and on the 
principle of statistical correctness permitting us to construct a 
correct multichannel Hamiltonian.

Within the framework of this approach many details can be changed and 
some improvements may be made. For example, instead of the Bonn potential 
we could opt for some other effective nucleon--nucleon interactions, 
we could include more kinds of bound states, and so on. Also, we have 
not discussed here the role of strangeness whose generation can be 
important as a diagnostic tool for the study of the quark--gluon plasma 
signals in realistic nuclear collisions [40]. The work on these 
problems is under progress. Changing particular details of the model 
might slightly alter some quantitative values, but the qualitative 
picture should remain the same. According to this picture, the 
deconfinement is a gradual crossover from hadron states to the 
quark--gluon plasma in agreement with the lattice numerical calculations 
[14].

\section{Discussion}

In this paper we suggested a general approach for treating complex 
systems whose constituent particles can form different bound states 
realized as compact clusters. The approach is based on three main 
notions: principle of statistical correctness, principle of cluster 
coexistence, and principle of potential scaling. The last two principles 
are compulsory for  a self--consistent treatment of clustering matter. 
The disregard of the first principle, that of statistical correctness, 
can sometimes be admissible when one is looking for a very rough 
approximate picture of clustering. However, if one wishes to get a really 
accurate description, the principle of statistical correctness must be 
satisfied. Moreover, neglecting this principle leads in some cases to the 
appearance of unphysical instabilities around phase transition points [28].

To illustrate the importance of using the princliple of statistical 
correctness, let us consider a simple clustering system consisting of two 
species, of unbound gluons and of glueballs that are bound gluon 
clusters. Include into consideration five lightest glueballs with the 
masses found in the bag--model calculations [41]. Take into account that 
the gluon as well as glueball chemical potentials are zero, 
$\;\mu_i=0\;$. Consider, for concreteness, the system with the 
$\;SU(2)\;$ symmetry when $\;N_c=2\;$ and the gluon degeneracy factor is 
$\;\zeta_g=6\;$. And let us calculate thermodynamic characteristics of 
such a system for two cases: when the principle of statistical correctness 
is satisfied and when it does not. In the second case we simply put zero 
the correcting term $\;CV\;$ in the Hamiltonian (11). The results of 
calculations are presented in figs.16-21, where the specific heat, 
pressure, and energy density are given in relative units, with respect to 
the corresponding quantities of the quarkless Stefan--Boltzmann plasma. 
The relative energy density and pressure are compared with the lattice 
numerical data, fitting the parameters of the model so that to have 
deconfinement at $\;T_d\cong 220\; MeV\;$. The qualitative behaviour of 
thermodynamic characteristics is similar in both cases, in the presence 
of the correcting term $\;CV\;$ and without it. At low temperatures, 
practically all gluons are clustered into glueballs. The glueball channel 
probability $\;w_G\;$ drastically drops down at the deconfinement 
temperature (fig.16). The gluon channel probability $\;w_g\;$ behaves 
oppositely to $\;w_G\;$, as is shown in fig.17. The deconfinement is a 
second order transition with a sharp peak of specific heat (fig.18). The 
comparison of the relative energy density with lattice numerical data for 
the case of the model without the correcting term is given in fig.19, as 
compared with the data of ref.[42], and in fig.20, as compared with those 
from ref.[43]. The relative energy density and pressure for the model 
satisfying the principle of statistical correctness are presented in fig.21,
compared with the lattice numerical data [43]. As is clearly seen from 
these comparisons, the model without the correcting term does not fit 
well the numerical data, while the statistically correct model is in
perfect agreement with the latter.

\vspace{1cm}

{\large{\bf Appendix. High--Temperature Limit}}

\vspace{3mm}

In section 5 we considered the high--temperature behaviour of pressure 
stating that it coincides with the high--temperature asymptotic expression 
for pressure in quantum chromodymanics. Here we show this explicitly.

The effective coupling parameter $\;g=g(T)\;$ in quantum chromodynamics 
is given by the equation
$$ g^2(T)\simeq\frac{24\pi^2}{(11N_c-2N_f)\ln(T/\Lambda)} , $$
where $\;N_f\;$ is the number of flavours and $\;\Lambda\;$ is a scaling 
parameter, $\;\Lambda\approx 200\; MeV\;$. The coupling $\;g(T)\ra 0\;$ 
as $\;T\ra\infty\;$. Therefore, perturbation theory in powers of $\;g\;$ 
becomes admissible. Perturbative expansions for the free energy have been 
obtained [44,45] to order $\;g^5\;$. The fourth and fifth orders of these 
expansions contain an arbitrary renormalization scale. The $\;QCD\;$ 
pressure can be written [17] as
$$ p =AT^4 , $$
where the factor $\;A\;$, as $\;T\ra\infty\;$, is
$$ A\simeq A_0+A_2g^2 +A_3g^3 , $$
with the coefficients
$$ A_0 =\frac{\pi^2}{45}\left [ N_c^2 - 1 +\frac{7}{4}N_cN_f + 15N_cN_f
\frac{\mu^2}{2\pi^2T^2}\left ( 1+\frac{\mu^2}{2\pi^2T^2}\right )\right ] , $$
$$ A_2 =-\frac{N_c^2-1}{144}
\left [ N_c+\frac{5}{4}N_f +9N_f\frac{\mu^2}{2\pi^2T^2}
\left ( 1 +\frac{\mu^2}{2\pi^2T^2}\right )\right ] , $$
$$ A_3 =\frac{N_c^2-1}{12\pi}\left ( \frac{1}{3}N_c +\frac{1}{6}N_f 
+\frac{1}{3}N_f\frac{\mu^2}{2\pi^2T^2}\right )^{3/2} ; $$
the chemical potentials of quarks are assumed to be equal to $\;\mu\;$.
When $\;g\ra 0\;$, only the term $\;A_0\;$ remains. The pressure 
$\;p\simeq A_0T^4\;$, with the relations in (45), coincides with 
expression (43).

In the particular case of $\;N_c=3\;$, and taking into account that 
$\;\mu\ra 0\;$ as $\;T\ra\infty\;$, we have
$$ A_0 =\frac{8\pi^2}{45}\left ( 1 +\frac{21}{32}N_f\right ) , $$
$$ A_2 =-\frac{1}{6}\left ( 1 +\frac{5}{12}N_f\right ) , $$
$$ A_3 =\frac{2}{3\pi}\left ( 1 +\frac{1}{6}N_f\right )^{3/2} . $$
Comparing the pressure $\;p=AT^4\;$ with the Stphan--Boltzmann limit from 
the end of section 5, we get
$$ \frac{p}{p_{SB}}\simeq 1+a_2g^2 +a_3g^3 $$
where
$$ a_2\equiv\frac{A_2}{A_0} , \qquad a_3\equiv\frac{A_3}{A_0} . $$
Thus, as $\;g\ra 0\;$, the pressure tends to the Stephan--Boltzmann form.

\newpage

\begin{center}
{\bf Figure Captions}
\end{center}

\vspace{5mm}

{\bf Fig.1} 

The probability of the  quark--gluon plasma channel as a 
function of temperature $\;\Theta=k_BT\;$ in $\;MeV\;$ and of the relative 
baryon density $\;n_B/n_{0B}\;$.

\vspace{5mm}

{\bf Fig.2} 

The $\;\pi\;$--meson channel probability on the 
temperature--baryon density plane.

\vspace{5mm}

{\bf Fig.3} 

The summary probability of the $\;\eta\;$--, $\;\rho\;$--, 
and $\;\om\;$--meson channels.

\vspace{5mm}

{\bf Fig.4} 

The nucleon channel probability.

\vspace{5mm}

{\bf Fig.5} 

The six--quark channel probability.

\vspace{5mm}

{\bf Fig.6} 

The channel probability of six--quarks in the Bose--condensed 
state.

\vspace{5mm}

{\bf Fig.7} 

The quark--gluon plasma channel probability at zero temperature 
as a function of the relative baryon density.

\vspace{5mm}

{\bf Fig.8} 

The nucleon channel probability at zero temperature vs. 
the relative baryon density.

\vspace{5mm}

{\bf Fig.9} 

The six--quark channel probability at zero temperature vs. 
the relative baryon density.

\vspace{5mm}

{\bf Fig.10} 

The pressure (in units of $\;J^4\;$) of the multichannel model 
on the temperature--baryon density plane.

\vspace{5mm}

{\bf Fig.11} 

The energy density (in units of $\;J^4\;$) of the multichannel
model.

\vspace{5mm}

{\bf Fig.12} 

The pressure--to--energy density ratio of the multichannel
model.

\vspace{5mm}

{\bf Fig.13} 

The specific heat (in units of $\;J^3\;$) for the multichannel
model.

\vspace{5mm}

{\bf Fig.14} 

The reduced specific heat of the multichannel model.

\vspace{5mm}

{\bf Fig.15} 

The compression modulus (in units of $\;J^4\;$) of the
multichannel model.

\vspace{5mm}

{\bf Fig.16}

The glueball channel probability as a function of temperature.

\vspace{5mm}

{\bf Fig.17}

Comparison of the glueball and gluon channel probabilities.

\vspace{5mm}

{\bf Fig.18}

Reduced specific heat for the gluon--glueball mixture.

\vspace{5mm}

{\bf Fig.19}

Relative energy density for the gluon--glueball model without the 
correcting term (solid line), as compared with the lattice Monte Carlo 
data [42].

\vspace{5mm}

{\bf Fig.20}

Relative energy density for the gluon--glueball model without the 
correcting term (solid line) compared with the lattice numerical 
simulations [43].

\vspace{5mm}

{\bf Fig.21}

Relative energy density and pressure for the gluon--glueball model satisfying 
the principle of statistical correctness (solid line) as compared with 
the lattice numerical data [43].

\newpage

\begin{center}
{\bf Table Caption}
\end{center}

\vspace{1cm}

The parameters of the particles included into the numerical investigation 
of thermodynamics of the multichannel model.

\end{document}